\documentclass[twocolumn,letter]{jpsj3} 
\title{Collapse of Ferrimagnetism 
in Two-Dimensional Heisenberg Antiferromagnet due to Frustration}
\catcode`\@=11
\def\simle{\mathrel{\mathpalette\@versim<}}   
\def\simge{\mathrel{\mathpalette\@versim>}}   
\def\@versim#1#2{\lower2.5pt\vbox{\baselineskip0pt \lineskip-.5pt
   \ialign{$\m@th#1\hfil##\hfil$\crcr#2\crcr\sim\crcr}}}
\catcode`\@=12

\author{Hiroki \textsc{Nakano}
\thanks{E-mail address: hnakano@sci.u-hyogo.ac.jp}, 
Tokuro \textsc{Shimokawa}, and 
Toru \textsc{Sakai}$^{1}$
}

\inst{Graduate School of Material Science, University of Hyogo,
Kouto 3-2-1, Kamigori, Ako-gun, Hyogo 678-1297, Japan \\
$^{1}$
Japan Atomic Energy Agency, SPring-8, 
Kouto 1-1-1, Sayo, Hyogo 679-5148, Japan
}

\recdate{\today}

\abst{We study ferrimagnetism in the ground state 
of the antiferromagnetic Heisenberg model 
on the spatially anisotropic kagome lattice, 
in which ferrimagnetism of the conventional Lieb-Mattis type 
appears in the region of weak frustration 
whereas the ground state is nonmagnetic 
in the isotropic case. 
Numerical diagonalizations of small finite-size clusters 
are carried out to examine the spontaneous magnetization. 
We find that the spontaneous magnetization changes continuously 
in the intermediate region 
between conventional ferrimagnetism 
and the nonmagnetic phase. 
Local magnetization of the intermediate state shows 
strong dependence on the site position, 
which suggests non-Lieb-Mattis ferrimagnetism. 
}

\kword{antiferromagnetic Heisenberg spin model,
ferrimagnetism, frustration, 
numerical-diagonalization method, Lanczos method}

\begin{document}
\maketitle


Ferrimagnetism has been studied extensively as 
an important phenomenon that has both 
ferromagnetic nature and antiferromagnetic nature 
at the same time. 
One of the fundamental keys to understanding ferrimagnetism 
is the Marshall-Lieb-Mattis (MLM) 
theorem\cite{Marshall,Lieb_Mattis}. 
This theorem clarifies some of the magnetic properties 
in the ground state of a system 
when the system has a bipartite lattice structure and 
when a spin on one sublattice interacts antiferromagnetically 
with a spin on the other sublattice.  
Under the condition that 
the sum of the spin amplitudes of spins in each sublattice 
is different between the two sublattices, one finds that 
the ground state of such a system exhibits ferrimagnetism. 
In this ferrimagnetic ground state, 
spontaneous magnetization is realized and 
its magnitude is a simple fraction 
of the saturated magnetization. 
We hereafter call ferrimagnetism of this type 
the Lieb-Mattis (LM) type. 

Some studies in recent years, on the other hand, reported cases 
when the magnitude of the spontaneous magnetization 
of the ferrimagnetism is not a simple fraction 
of the saturated magnetization\cite{Ivanov_Richter,
SYoshikawa_Miya,Hida,Hida_Takano,Montenegro,
Hida_Takano_Suzuki,Shimokawa}. 
The ferrimagnetic ground state of this type 
is a nontrivial quantum state whose behavior 
is difficult to explain well only within the classical picture. 
Ferrimagnetism of this type 
was first predicted in ref.~\ref{Sachdev_Senthil} 
using the quantum rotor model. 
The mechanism of this ferrimagnetism has not been 
understood sufficiently up to now.  
Hereafter, we call this case the non-Lieb-Mattis (NLM) type. 
Note that in the cases when NLM ferrimagnetism is present, 
the structure of the lattices is limited to being 
one-dimensional. 
Recall that the above conditions of the MLM theorem 
do not include the spatial dimension of the system;  
the MLM theorem holds irrespective of the spatial dimensionality. 
We are then faced with a question: 
can NLM ferrimagnetism be realized 
when the spatial dimension is more than one? 

The purpose of this letter is to answer the above question 
concerning the existence of NLM ferrimagnetism 
in higher dimensions. 
In this letter, we consider a case when we introduce 
a frustrating interaction into a two-dimensional lattice 
whose interactions satisfy the conditions of the MLM theorem. 
When the frustrating interaction is small, 
ferrimagnetism of the LM type survives; however, 
the ferrimagnetism is destroyed with the increase 
in the frustrating interaction and 
the system finally becomes nonmagnetic due to 
the considerably large frustrating interaction. 
We examine the behavior of the collapse of the ferrimagnetism 
and the existence of an intermediate region 
between the LM ferrimagnetic and nonmagnetic phases 
by means of the numerical-diagonalization method 
applied to finite-size clusters.  
Our study of the two-dimensional system successfully 
clarifies the existence of the intermediate phase 
and captures a feature of NLM ferrimagnetism. 

\begin{figure}[htb]
\begin{center}
\includegraphics[width=6cm]{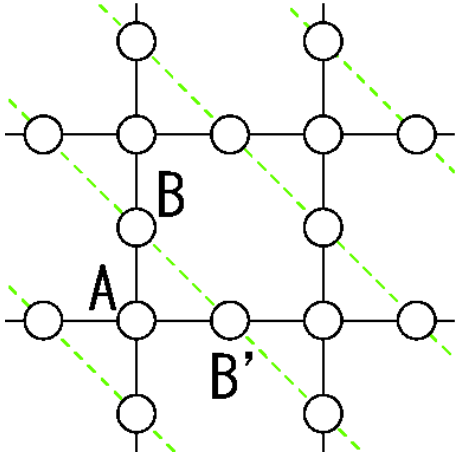}
\end{center}
\caption{Network of interactions in the system 
and sublattices A, B, and B$^{\prime}$. 
Black straight lines and green dotted lines 
denote interactions of $J_1$ and $J_2$, 
respectively. 
Open circles at lattice points represent $S=1/2$ spins. 
The system of classical spins in this lattice 
was studied by Monte Carlo simulations 
in ref.~\ref{Kaneko_Imada}. 
}
\label{fig1}
\end{figure}

First, we explain the model Hamiltonian 
examined in this letter. 
The Hamiltonian is given by 
\begin{eqnarray}
{\cal H} &=& \sum_{i\in {\rm A},j\in {\rm B}} 
J_{1} \mbox{\boldmath $S$}_{i} \cdot \mbox{\boldmath $S$}_{j} 
+
\sum_{i\in {\rm A},j\in {\rm B}^{\prime}} 
J_{1} \mbox{\boldmath $S$}_{i} \cdot \mbox{\boldmath $S$}_{j} 
\nonumber \\
& & 
+
\sum_{i\in {\rm B},j\in {\rm B}^{\prime}} 
J_{2} \mbox{\boldmath $S$}_{i} \cdot \mbox{\boldmath $S$}_{j}, 
\label{Hamiltonian}
\end{eqnarray}
where $\mbox{\boldmath $S$}_{i}$ denotes 
an $S=1/2$ spin operator at site $i$. 
Sublattices A, B, and B$^{\prime}$ and the network 
of antiferromagnetic interactions $J_{1}$ and $J_{2}$ are 
depicted in Fig.~\ref{fig1}. 
Here, we consider the case of isotropic interactions. 
The system size is denoted by $N_{\rm s}$; 
the saturation magnetization is $M_{\rm sat}=N_{\rm s}/2$. 
Energies are measured in units of $J_{1}$; 
thus, we take $J_{1}=1$ hereafter. 
We examine the properties of this model 
in the range of $0 < J_{2}/J_{1} \le 1$. 
Note that in the case of $J_{2}=0$, 
sublattices B and B$^{\prime}$ are combined into 
a single sublattice; the system satisfies the above conditions 
of the MLM theorem. 
Thus, ferrimagnetism of the LM type is 
exactly realized in this case. 
In the case of $J_{2}=J_{1}$, on the other hand, 
the lattice of the system is reduced to the kagome lattice. 
The ground state of the system on the kagome lattice 
without a magnetic field is known to be singlet 
from numerical-diagonalization 
studies\cite{Lecheminant,Waldtmann,Hida_kagome,HN_Sakai2010}, 
which indicates that the ground state is nonmagnetic. 
One thus finds that 
LM ferrimagnetism collapses between $J_{2}=0$ and $J_{2}=J_{1}$. 
Consequently, we survey the region between the two cases. 

Next, we discuss the method we use here, which is  
numerical diagonalization based 
on the Lanczos algorithm. 
It is known that this method is nonbiased 
beyond any approximations and 
reliable for many-body problems such as the present model. 
A disadvantage of this method is that the available system sizes 
are limited to being small 
because the dimension of the matrix grows exponentially 
with respect to the system size. 
To treat systems that are as large as possible, 
we have developed parallelization 
in our numerical calculations using the OpenMP and MPI 
techniques, 
either separately or in a hybrid way\cite{comments2para}.  

\begin{figure}[htb]
\begin{center}
\includegraphics[width=7cm]{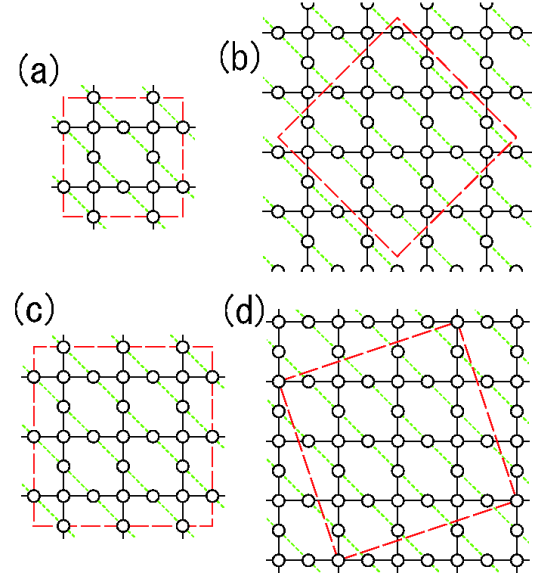}
\end{center}
\caption{Finite-size clusters: 
(a) $N_{\rm s}=12$, (b) $N_{\rm s}=24$, (c) $N_{\rm s}=27$, and 
(d) $N_{\rm s}=30$ under the periodic boundary condition, 
where red dashed lines denote a single finite-size cluster 
with each system size. 
Black straight lines and green dotted lines are the 
same as in Fig.~1. 
Note that cluster (c) under the open boundary condition 
includes $N_{\rm s}=33$ spins. 
}
\label{fig2}
\end{figure}

In this letter, we treat the finite-size clusters 
depicted in Fig.~\ref{fig2} when the system sizes are 
$N_{\rm s}=12$, $N_{\rm s}=24$, $N_{\rm s}=27$, and 
$N_{\rm s}=30$ under the periodic boundary condition 
and 
$N_{\rm s}=33$ under the open boundary condition. 
Note that each of these clusters forms a regular square 
although clusters (b) and (d) are tilted. 
The next larger size under the condition 
that a regular square is formed 
is $N_{\rm s}=48$, 
which is too large to handle using the present method, 
even when one uses modern supercomputers. 

\begin{figure}[htb]
\begin{center}
\includegraphics[width=2cm]{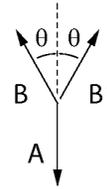}
\end{center}
\caption{Ferrimagnetic spin direction in the classical picture. 
}
\label{fig3}
\end{figure}

Before our numerical-diagonalization results 
for the finite-size clusters are presented, 
let us consider the directions of the spins 
in the ground state within the classical picture. 
We here examine the spin directions 
of classical vectors with length $S$ 
depicted in Fig.~\ref{fig3}. 
One obtains the energy of the spin state with angle $\theta$ 
to be $E/J_{1}=-(2N_{\rm s}/3)S^{2}[2\cos(\pi-\theta)
+(J_2/J_1)\cos(2\theta)]$. 
This expression of the energy indicates that for $J_2/J_1\le1/2$, 
the state of $\theta=0$, namely, ferrimagnetism of the LM type,  
is realized. 
Thus, the normalized magnetization of this state 
is $M/M_{\rm sat}=1/3$.  
One finds, on the other hand, that for $J_2/J_1>1/2$, 
the lowest-energy state is realized for nonzero $\theta$ 
when $J_1/J_2=2\cos(\theta)$ is satisfied. 
The normalized magnetization of this state 
is $M/M_{\rm sat}=(\frac{J_1}{J_2}-1)/3$.  
When $J_2/J_1$ becomes unity, the magnetization finally vanishes. 
This classical argument will be compared with 
our finite-size results obtained from numerical diagonalizations. 

\begin{figure}[htb]
\begin{center}
\includegraphics[width=7.5cm]{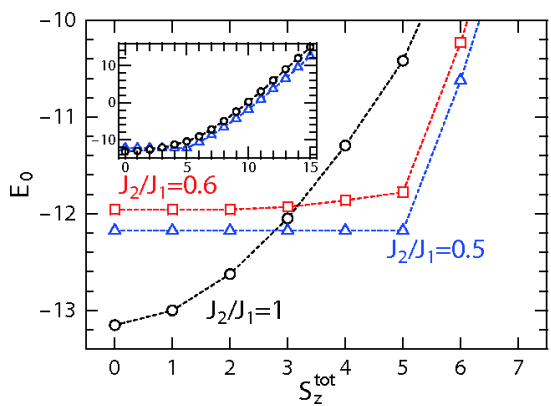}
\end{center}
\caption{Lowest energy in each subspace 
of $S_{z}^{\rm tot}$ for the system of $N_{\rm s}=30$ 
depicted in Fig.~\ref{fig2}(d). 
Results for $J_{2}/J_{1}=1$, 0.5, and 0.6 
are presented by black circles, blue triangles, and 
red squares, respectively. 
Inset: our data in the entire range of $S_{z}^{\rm tot}$ 
given for $J_{2}/J_{1}=1$ and 0.5. 
}
\label{fig4}
\end{figure}

Now, let us present our numerical results for the quantum case. 
First, we show our data for the lowest energy 
in each subspace of $S_{z}^{\rm tot}$, 
which reveal the magnetization of the systems. 
Figure \ref{fig4} depicts our results 
for the system with $N_{\rm s}=30$ 
depicted in Fig.~\ref{fig2}(d). 
Note that $M_{\rm sat}=15$ in this case. 
For $J_2/J_1=0.5$, the energies from $S_{z}^{\rm tot}=0$
to $S_{z}^{\rm tot}=5$ are numerically identical, 
which means that $M/M_{\rm sat}$ becomes 1/3 and that 
ferrimagnetism of the LM type is realized. 
For $J_2/J_1=1$, the energy for $S_{z}^{\rm tot}=0$ 
is lower than the other energies for larger $S_{z}^{\rm tot}$.  
The ground state of this case is nonmagnetic. 
For $J_2/J_1=0.6$, the energies from $S_{z}^{\rm tot}=0$
to $S_{z}^{\rm tot}=2$ are the same; thus, we find that 
the spontaneous magnetization is $M=2$, 
which is smaller than the value 
for ferrimagnetism of the LM type. 
One finds that a state with intermediate magnetization 
appears between LM-type ferrimagnetism and 
the nonmagnetic state, 
at least according to the finite-size calculations.  

\begin{figure}[htb]
\begin{center}
\includegraphics[width=7.5cm]{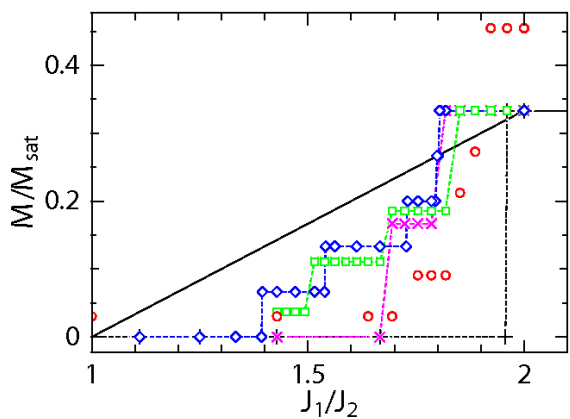}
\end{center}
\caption{Dependence of the spontaneous magnetization 
normalized by the saturated magnetization 
on $J_{1}/J_{2}$. 
The solid line represents the result for the magnetization 
within the classical picture shown in Fig.~\ref{fig3}. 
Note that we take the $J_{1}/J_{2}$ dependence 
as the abscissa 
because the classical magnetization shows 
linear dependence not on $J_{2}/J_{1}$ 
but on $J_{1}/J_{2}$. 
Results for $N_{\rm s}=12$, 24, 27, and 30 
under the periodic boundary condition are presented 
by black pluses, violet crosses, green squares, 
and blue diamonds, respectively. 
Red circles denote results for 
$N_{\rm s}=33$ under the open boundary condition. 
}
\label{fig5}
\end{figure}

Next, we examine the region of such an intermediate state 
for various system sizes; 
our results are depicted in Fig.~\ref{fig5}. 
In the case of $N_{\rm s}=12$, 
the intermediate state 
between LM-type ferrimagnetism and the nonmagnetic state 
is absent; 
on the other hand, the intermediate region exists 
for all the larger systems. 
Note that 
the width of the intermediate region increases 
for the cases under the periodic boundary condition 
when $N_{\rm s}$ is increased. 
This indicates that the intermediate phase 
is present in the thermodynamic limit. 
One of the characteristics observed 
is that the continuity of the magnetization 
improves with increasing $N_{\rm s}$. 
In the cases under the open boundary condition, 
we successfully detect the intermediate phase 
although its width is relatively smaller. 
The width for the $N_{\rm s}=33$ case 
under the open boundary condition is close 
to that for the $N_{\rm s}=24$ case 
under the periodic boundary condition. 
This is consistent with the fact that 
there are 21 sites in the inner part of cluster (c) of 
$N_{\rm s}=33$ under the open boundary condition. 
Our present results for both boundary conditions 
imply that 
the presence of the intermediate phase is irrespective 
of the boundary conditions. 

\begin{figure}[htb]
\begin{center}
\includegraphics[width=7.5cm]{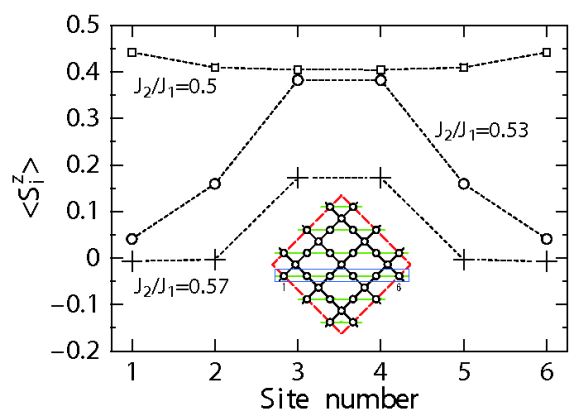}
\end{center}
\caption{Local magnetizations of the cluster with $N_{\rm s}=33$ 
under the open boundary condition. 
Results for $J_{2}/J_{1}=0.5$, 0.53, and 0.57 are presented 
for the sites surrounded by the blue rectangles in the inset. 
Site numbers 1 to 6 
correspond to the sites from left to right in the blue rectangle. 
}
\label{fig6}
\end{figure}

An important characteristic of NLM ferrimagnetism 
is that the local magnetization in sublattice 
exhibits long-distance periodicity, which is absent 
in LM-type ferrimagnetism. 
Note that one cannot detect this periodicity in the cases 
under the periodic boundary condition. 
We thus examine the local magnetization 
in the intermediate phase for the case 
under the open boundary condition; 
the results for $N_{\rm s}=33$ are depicted in Fig.~\ref{fig6}. 
For $J_2/J_1=0.5$ with LM-type ferrimagnetism, 
the local magnetization shows weak dependence on the position 
of sites, although $\langle S_i^z \rangle$ 
at edge sites 1 and 6 
is slightly larger than those at interior sites,
where the site numbers are illustrated 
in the inset of Fig.~\ref{fig6}. 
This small difference originates from the edge effect 
due to the open boundary condition. 
For $J_2/J_1=0.5$, the edge effect does not seem to affect 
$\langle S_i^z \rangle$ at internal sites. 
For $J_2/J_1=0.53$ and 0.57, on the other hand, 
$\langle S_i^z \rangle$ at edge sites 1 and 6 becomes 
very small. 
The behavior of this appearance of the edge effect 
is different from the case of $J_2/J_1=0.5$. 
For $J_2/J_1=0.53$ and 0.57,  
one finds a strong dependence of $\langle S_i^z \rangle$ 
on the position of the site from site 2 to site 5. 
For $J_2/J_1=0.53$, 
$\langle S_i^z \rangle$ at sites next to the edges 
seems to be affected by the edge sites. 
It is unclear whether or not the case of $J_2/J_1=0.53$ 
corresponds to NLM type ferrimagnetism at present.  
For $J_2/J_1=0.57$, on the other hand, 
the strong dependence on the site position 
suggests the existence of origins 
that are different from the edge effect. 
The system size $N_{\rm s}=33$ is sufficiently small 
for long-distance periodicity to be observed clearly.  
Although the present results are not decisive evidence 
of the periodicity, our finding of the large change 
in $\langle S_i^z \rangle$ is considered as possible evidence. 
In order to obtain decisive evidence, calculations 
on systems of larger sizes are required, 
which are unfortunately difficult at the present time. 
Instead of the present two-dimensional kagome case, 
we are now examining a quasi-one-dimensional system 
on a kagome stripe lattice. 
Both systems partly share the same lattice structure.  
The system on the kagome stripe lattice reveals 
the clear appearance of NLM ferrimagnetism 
in the intermediate region\cite{Simokawa_kagome_stripe}. 
Results will be published elsewhere. 


The phenomenon of ground-state magnetization 
changing continuously with respect to a continuous parameter 
in a model Hamiltonian has been reported in other cases. 
Tonegawa and co-workers reported such a phenomenon 
in spin systems with anisotropic 
interactions\cite{Tone1,Tone2,Tone3,Tone4,Tone5}. 
It is unclear at present whether or not the states 
of this continuously changing magnetization 
in the anisotropic case show long-distance periodicity 
because the behavior of local magnetization 
has not been investigated yet. 
Since this phenomenon disappears in the isotropic case 
when the quantum effect is stronger 
than that in the anisotropic case, 
this phenomenon is considered to arise from the anisotropy. 
From this point of view, 
the origin of this phenomenon seems to be different from 
that of intermediate ferrimagnetism in the isotropic case 
studied here. 
Another reported phenomenon is partial ferromagnetism 
in the Hubbard model\cite{HN_YT2003,HN_YT2004} 
when the system is hole-doped 
near the half-filled Mott insulator. 
The origin of this phenomenon has been clarified 
to be the formation of spin polarons around doped holes. 
The mechanism of these two cases is different 
from the present case of NLM ferrimagnetism. 

Finally, we briefly discuss possible future experiments. 
For volborthite, 
eq.~(\ref{Hamiltonian}) was proposed as a model Hamiltonian 
from the argument 
of its crystal structure\cite{MALafontaine,ZHiroi}, 
although NLM ferrimagnetism has not yet been observed 
in this material. 
A theoretical study on the spatial anisotropy of this material 
indicated that the deviation of the anisotropy 
from the isotropic kagome point 
is not particularly large\cite{PSindzingre}. 
This is consistent with our present result 
because the nonmagnetic ground state is realized 
around the region of weak anisotropy 
as shown in Fig.~\ref{fig5}. 
In order to observe NLM ferrimagnetism experimentally, 
it is necessary to realize a case with larger anisotropy. 
The measurement of volborthite under high pressure 
in the direction of the $a$-axis or 
discoveries of new materials might lead to such an observation. 


In summary, we have clearly shown the existence 
of a ground state of non-Lieb-Mattis type ferrimagnetism 
in a two-dimensional lattice 
that lies between 
the well-known Lieb-Mattis type ferrimagnetic phase 
and the nonmagnetic 
phase including the kagome-lattice system. 
The nontrivial ferrimagnetism we have found 
in the intermediate phase occurs 
as a consequence of magnetic frustration. 
Our present result indicates that 
non-Lieb-Mattis ferrimagnetism is a general phenomenon 
irrespective of the spatial dimensionality. 

\section*{Acknowledgments}
We wish to thank 
Professor K.~Hida, Professor T.~Tonegawa, 
Professor S.~Miyashita, Professor M.~Imada,
and Dr.~Y.~Okamoto 
for fruitful discussions. 
This work was partly supported by a Grant-in-Aid (No.~20340096) 
from the Ministry of Education, Culture, Sports, Science 
and Technology of Japan. 
This work was partly supported by 
a Grant-in-Aid (No. 22014012) 
for Scientific Research and Priority Areas 
``Novel States of Matter Induced by Frustration'' 
from the Ministry of Education, Culture, Sports, Science 
and Technology of Japan. 
Nonhybrid thread-parallel calculations 
in the numerical diagonalizations were based 
on TITPACK ver.2, coded by H. Nishimori. 
Part of the computations were performed using 
the facilities of 
Information Technology Center, Nagoya University;  
Department of Simulation Science, 
National Institute for Fusion Science; 
and the Supercomputer Center, 
Institute for Solid State Physics, University of Tokyo. 



\begin{thebibliography}{99}
\bibitem{comments2para}
This hybrid code was originally developed 
in a study on the estimate of the Haldane gap\cite{HN_Terai}.
\end{thebibliography}
\end{document}